\documentstyle[11pt,newpasp,twoside,epsf]{article}
\begin{document}
\title{Photometry of pulsating stars in the Magellanic Clouds as
  observed in the MOA project} 

\author{J.\ B.\ Hearnshaw}
\affil{Dept. of Physics and Astronomy, University of Canterbury,
  Christchurch, New Zealand}

\author{I.\ A.\ Bond} 
\affil{Dept. of Physics, University of Auckland, New Zealand 
and Dept. Physics and Astronomy, University of Canterbury, New Zealand}

\author{N.\ J.\ Rattenbury} 
\affil{Dept. of Physics, University of Auckland, New Zealand}

\author{S.\ Noda}
\affil{Solar-Terrestrial Environment Laboratory, Nagoya University, 
Nagoya, Japan}

\author{M.\ Takeuti}
\affil{Tohoku University, Sendai, Japan}

\bigskip

\noindent \hspace*{3em} F.\ Abe\footnote{Solar-Terrestrial Environment
  Lab, Nagoya University, Nagoya, Japan}, 
B.\ S.\ Carter\footnote{Carter Observatory, Wellington, N.Z.}, 
R.\ J.\ Dodd$^{2}$,
M.\ Honda\footnote{Inst. Cosmic Ray Research, University of Tokyo, Japan}, 
J.\ Jugaku\footnote{Institute for Civilization, Tokai University, Japan}, 
S.\ Kabe\footnote{High Energy Accelerator Research Organization (KEK), 
Tsukuba, Japan}, \\ 
\noindent \hspace*{3em} P.\ M.\ Kilmartin\footnote{Dept.  of  Physics \& 
  Astronomy, University  of  Canterbury, N.Z.} 
\footnote{Dept. of Physics, University of Auckland, N.Z.},
B.\ S.\ Koribalski\footnote{Australia Telescope National Facility,
  Epping, Australia}, 
Y.\ Matsubara$^{1}$, 
K.\ Masuda$^{1}$, \\ 
\noindent \hspace*{3em} Y.\ Muraki$^{1}$, 
T.\ Nakamura\footnote{Research Inst. Fundamental Physics, Kyoto
  University, Japan},
G.\ R.\ Nankivell\footnote{Lower Hutt, N.Z.}, 
M.\ Reid\footnote{Dept. of Physics, Victoria University Wellington, 
N.Z.}, \\ 
\noindent \hspace*{3em} N.\ J.\ Rumsey$^{10}$, 
To.\ Saito\footnote{Tokyo Metropolitan College of Aeronautics, Japan}, 
H.\ Sato$^{9}$, 
M.\ Sekiguchi$^{3}$,
D.\ J.\ Sullivan$^{11}$, \\ 
\noindent \hspace*{3em} T.\ Sumi$^{1}$, 
Y.\ Watase$^{5}$,
T.\ Yanagisawa$^{1}$, 
P.\ C.\ M.\ Yock$^{7}$ 
and  \\ 
\noindent \hspace*{3em} M.\ Yoshizawa\footnote{National Astronomical
  Observatory, Mitaka, Japan}

\newpage
\begin{abstract}
A review of the MOA (Microlensing Observations in Astrophysics) project
is presented. MOA is a collaboration of approximately 30 astronomers
from New Zealand and Japan established with the aim of finding and
detecting microlensing events towards the Magellanic Clouds and the
Galactic bulge, which may be indicative of either dark matter or of
planetary companions. The observing program commenced in 1995, using
very wide band blue and red filters and a nine-chip mosaic CCD camera.

As a by-product of these observations a large database of CCD photometry
for 1.4 million stars towards both LMC and SMC has been established. In
one preliminary analysis 576 bright variable stars were confirmed,
nearly half of them being Cepheids. Another analysis has identified
large numbers of blue variables, and 205 eclipsing binaries are included
in this sample. In addition 351 red variables (AGB stars) have been
found. Light curves have been obtained for all these stars. The
observations are carried out on a 61-cm f/6.25 telescope at Mt John
University Observatory where a new larger CCD camera was installed in
1998 July. From this latitude ($44^{\circ}$ S) the Magellanic Clouds can be
monitored throughout the year.

\end{abstract}

\section{Introduction}

The MOA project  {\bf M}icrolensing {\bf O}bservations in {\bf A}strophysics
is a collaboration of about 30 astronomers/physicists in New Zealand
and Japan working in about 10 different institutions (headquarters Nagoya
  and Auckland universities).
The project was established in 1995.
The MOA project goals are
(a) the search for dark matter by microlensing in the galactic halo, 
(b) the search for planets by microlensing, in halo and galactic bulge, 
(c) the study of variable stars in the Magellanic Clouds and bulge and
(d) the optical identification of $\gamma$-ray bursters (GRBs)

The observational program is undertaken on the 61-cm Boller \& Chivens
telescope at the University of Canterbury's Mt John University
Observatory in the centre of the South Island of New Zealand. Here there
is a dark sky background and typically 1--3 arcsec seeing; between a
third and a half of the weather is suitable for CCD photometry. At the
latitude of $44^{\circ}$ S, the Magellanic Clouds are circumpolar and
can be observed all year.

The telescope is a Ritchey-Chr\'etien cassegrain instrument modified to
accommodate wide-field f/6.25 optics and a modern computer control system.

\section{Observing program}

Three series of MOA observations have been undertaken as follows
\begin{enumerate}
\item Jan 1996 -- Dec 1996 f/13.5 Cass.\ $30^{\prime} \times
  30^{\prime}$, MOAcamI
\item Jan 1997 -- Aug 1998 f/6.25 Cass.\ $1^{\circ} \times 1^{\circ}$, MOAcamI
\item Aug 1998 -- present f/6.25 Cass.\ $0.92^{\circ} \times
  1.39^{\circ}$, MOAcamII
\end{enumerate}
In all cases, very broad-band $B$ and $R$ filters have been used,
covering respectively about 395--620 nm ($\lambda_{\rm eff} \simeq 500$
nm) and 620--1050 nm ($\lambda_{\rm eff} \simeq 700$ nm) (see Fig. 1).
Series 1 used the f/13.5 optics and encompassed $\simeq 3 \times 10^{5}$
stars in three LMC fields. Series 2 covered three fields (nlmc1,2,3) in
the LMC bar ($1 \times 10^{6}$ stars) plus two SMC fields (smc1,2) ($4
\times 10^{5}$ stars).

\begin{figure}[h]
\plotfiddle{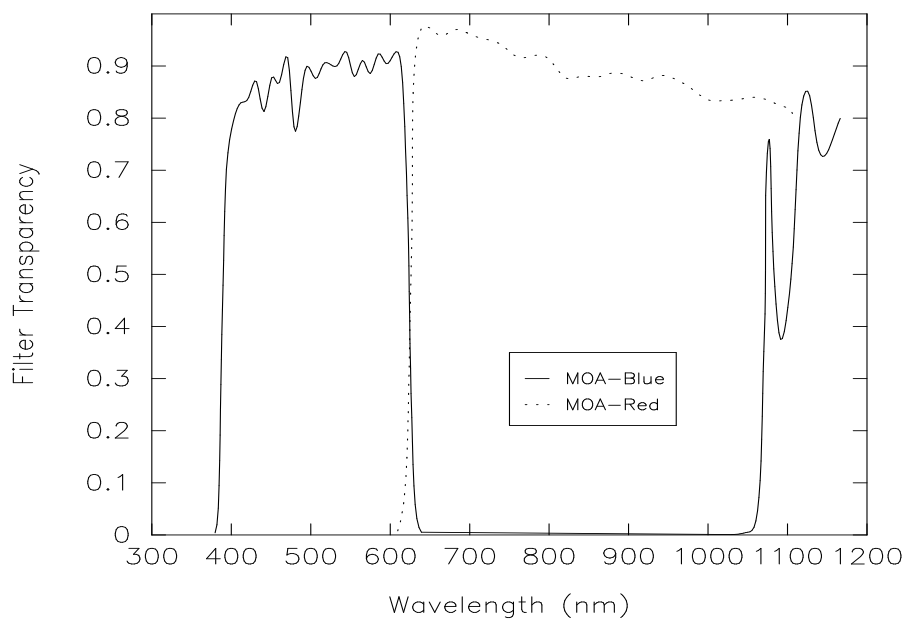}{9cm}{0}{80}{65}{-230}{-120}

\caption{MOA filter functions in $B$ and $R$ broad bands}
\end{figure}

For series 3, which is currently in progress, the new MOAcamII is 
used  to observe 16 LMC fields (all year) ($\sim 3 \times 10^{6}$ stars) 
plus 8 SMC fields (all year) ($\sim 1 \times 10^{6}$ stars) 
plus 16 galactic bulge fields (winter only)($\sim 5 \times 10^{6}$
stars). Each of the MOAcamII fields covers about 1.25 square degrees. 
Exposure times are 5 min.\ in $B$ and $R$ in the Clouds, 3 minutes in the 
Galactic bulge.

The CCD camera MOAcamI comprises a mosaic of nine TI TC215 $1000 \times
1018$ pixel CCDs whose pixel size is 12 $\mu$m ($\equiv
0.3^{\prime\prime}$ at f/13.5, $0.65^{\prime\prime}$ at f/6.25). The
chips are not butted, so four consecutive interleaved exposures are
required to cover a $1^{\circ} \times 1^{\circ}$ field.

The recently acquired MOAcamII has now replaced the first camera. It
consists of three butted SITe $2048 \times 4096$-pixel thinned, 
back-illuminated CCDs  with 
15-$\mu$m pixels ($\equiv 0.81^{\prime\prime}$ at f/6.25). The
peak QE is 85\%. The area of the CCD surface is $6.1 \times 9.2$ cm, 
and there are 24 Mpixels. Both cameras operated at a temperature of 
$-100^{\circ}$ C with liquid N$_{2}$.

Generally we expose each Magellanic Cloud field once or sometimes twice
per night in both $B$ and $R$, and each Galactic bulge field about four times
per night in $R$ and once in $B$, unless a known high magnification
microlensing event is in progress, in which case frequent successive
exposures are made.

\section{Data reduction}

MOA observations of CCD images are recorded to 8-mm Exabyte tape for
sending to Japan, while an archival copy on DLT tape is retained at Mt
John. For series 1 and 2 observations, all the reductions have been
completed using the DoPhot software package. One of us (I.B.) completed
an archived database of all series 1 and 2 frames giving magnitudes in $B$ 
and $R$. In the case of series 2 images, there are some 200 data points
in each colour and about $1.4 \times 10^{6}$ stars.

Series 3 images using MOAcamII have hitherto not been reduced. I.\ Bond
is currently perfecting implementation of the Alard \& Lupton (1997)
difference imaging technique, which already has shown that a substantial
improvement in photometric precision is possible for crowded fields in
only moderate seeing. It is hoped to reduce series 3 images early in
2000 and then to be able to reduce all images on site at the observatory 
in nearly real time. A Sun Enterprise 450 computer with 90 Gbyte hard
disk drive is used for on-site data analysis at Mt John.

Mt John $B,R$ MOA photometry gives colours which transform linearly to
Johnson colours, as calibrated by Reid, Dodd \& Sullivan (1997), as
follows:
\[ (B-V)_{\rm J} = 0.356 + 1.036(B-R)_{\rm MOA} \]
\[ (R-I)_{\rm J} = 0.150 + 0.612(B-R)_{\rm MOA} \]
The calibration of MOA magnitudes gives the following transformation
from the MOA instrumental system to that of the HST Guide Star Catalogue.
\[ m_{R}({\rm GSC}) = m_{R}({\rm MOA}) +25.10\]
\[ m_{B}({\rm GSC}) = m_{B}({\rm MOA}) +24.82 \]

The typical precision of the series 2 MOA magnitudes is $\pm 0.13$ mag. 
at $B$ or $R \sim 19$; $\pm 0.04$ mag.\ at $B$ or $R \sim 14$ -- 15.

\section{Variable stars in the MOA database}


\begin{figure}[t]
\plotfiddle{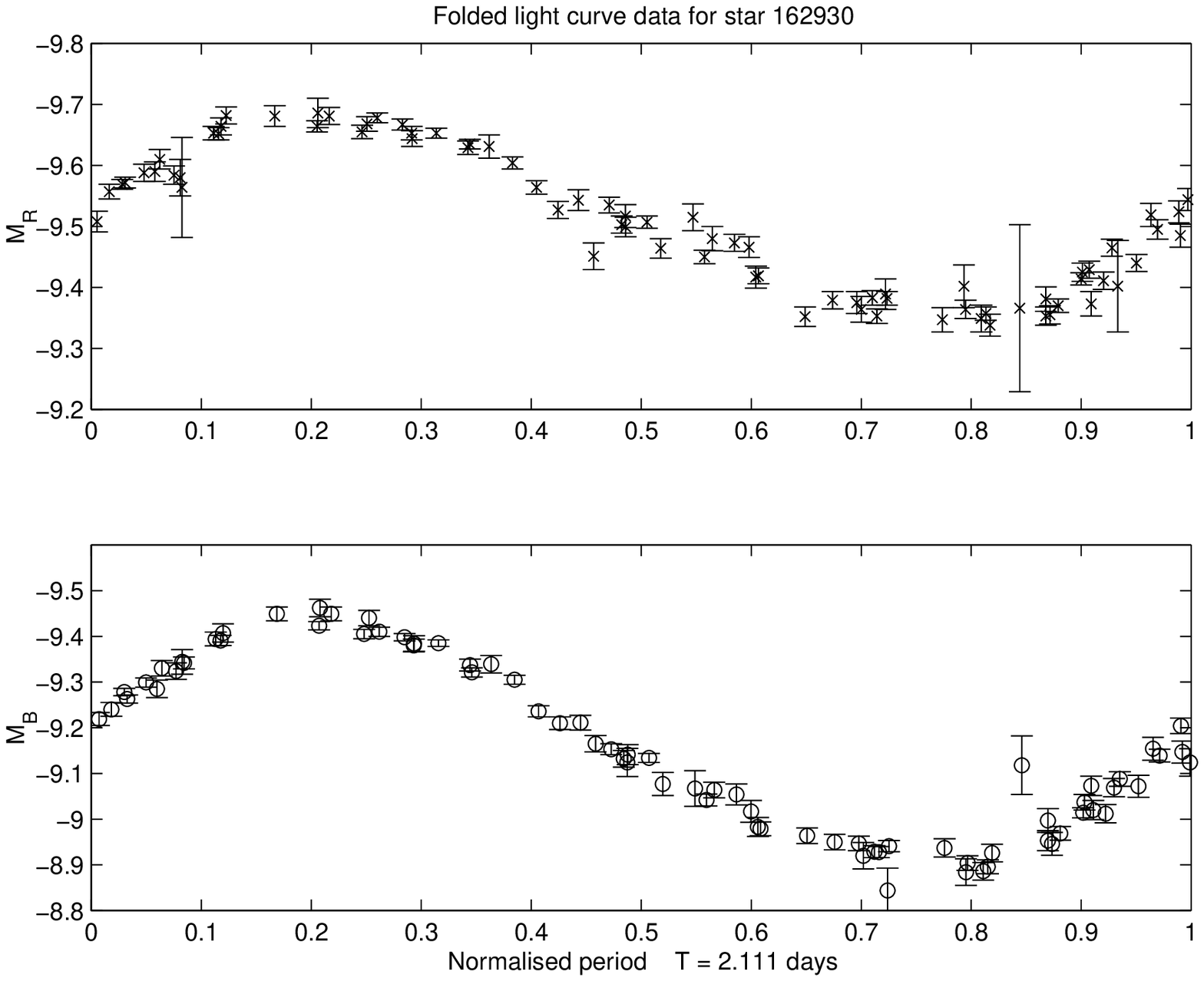}{9cm}{0}{40}{40}{-230}{30}
\plotfiddle{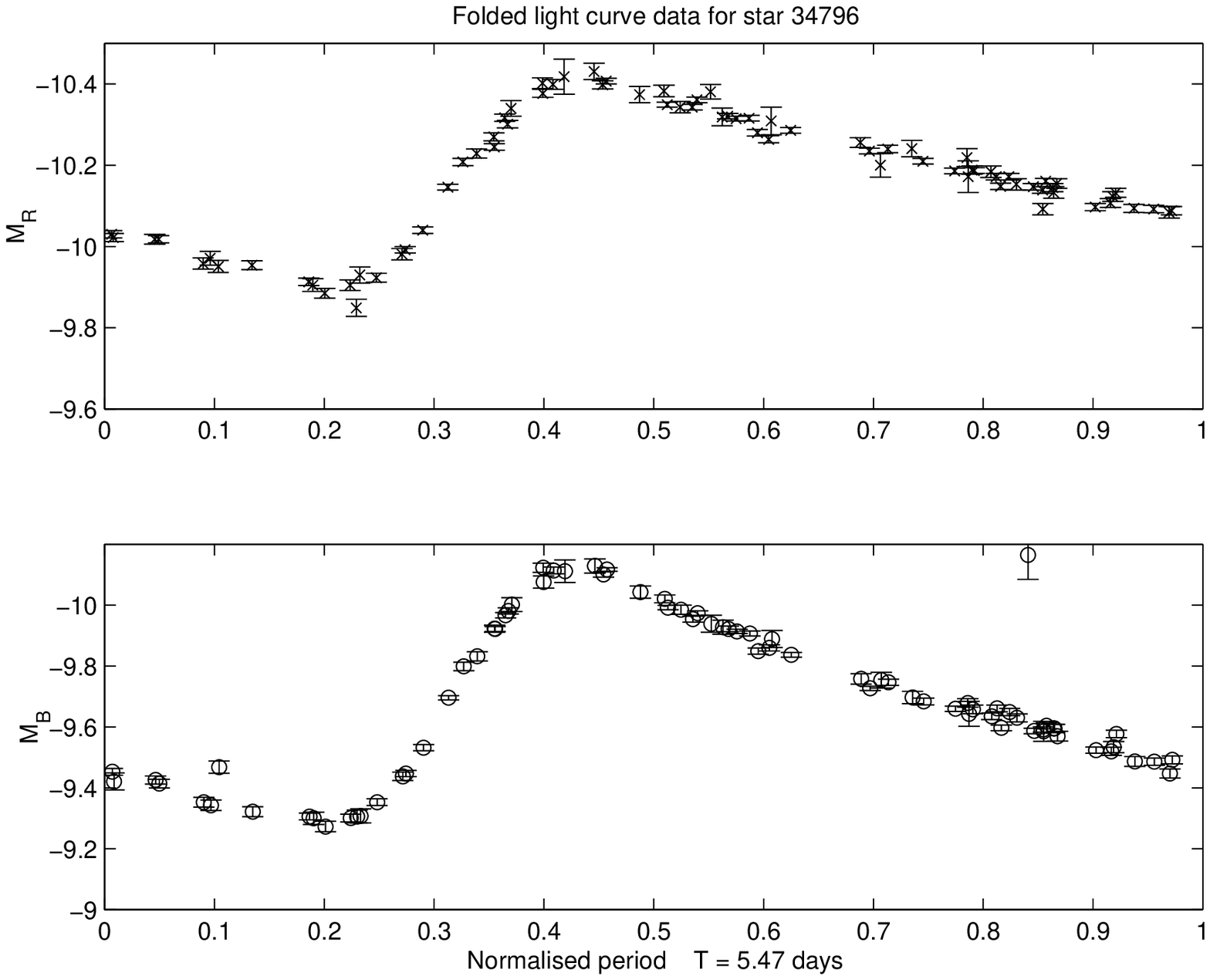}{0cm}{0}{40}{40}{-30}{55}
\plotfiddle{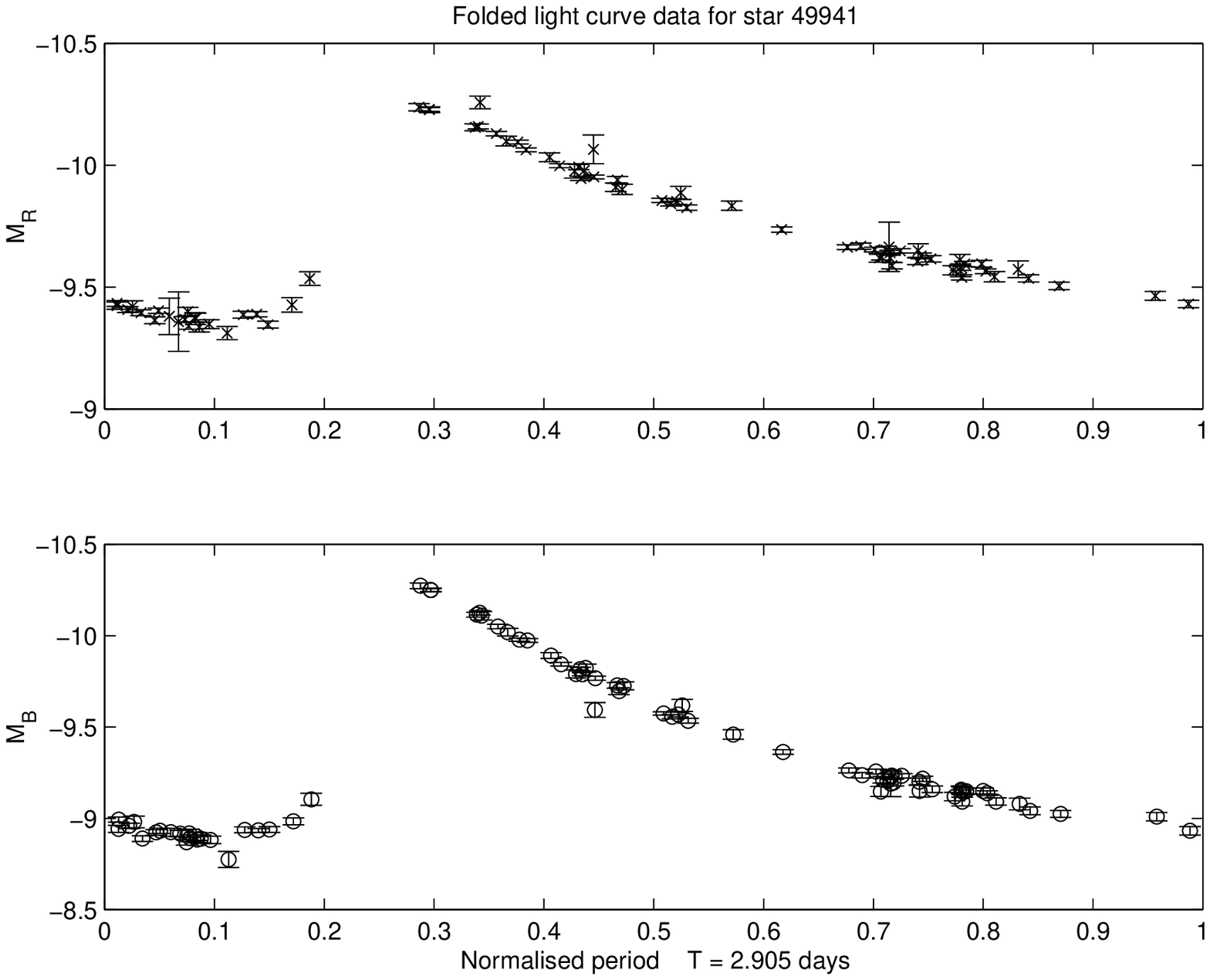}{0cm}{0}{40}{40}{-230}{-85}
\plotfiddle{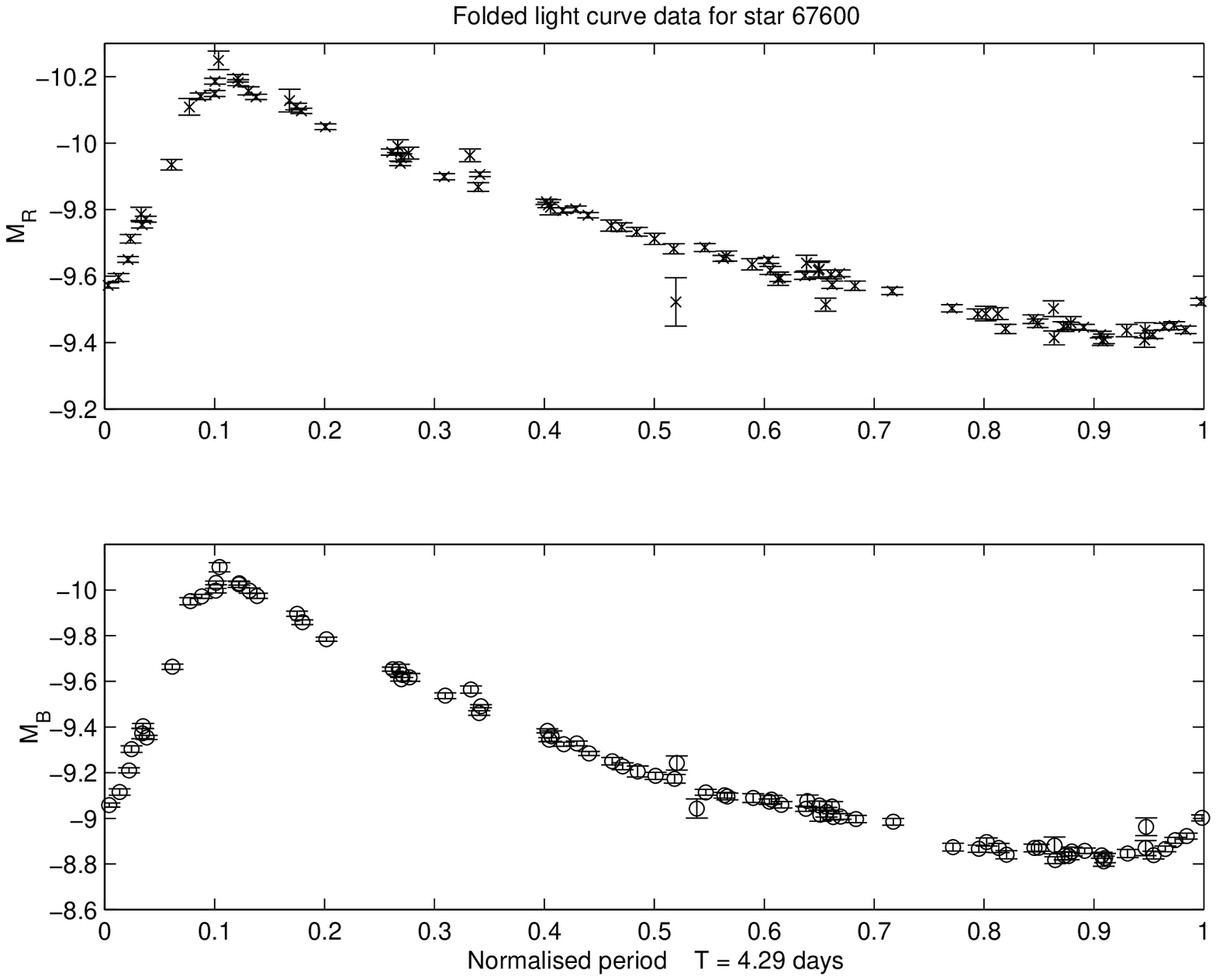}{0cm}{0}{40}{40}{-30}{-60}




\caption{$B$ and $R$ phased light curves for four Cepheids in field smc2}
\end{figure}

Three surveys on variable stars in the MOA database have been undertaken
or are in progress. One of these has been by S.\ Noda (Nagoya Univ., MSc)
and M.\ Takeuti (Tohoku Univ., Sendai), and is a study of variables in
series 2 observations. The stars are selected by several criteria (see Noda et
 al. in these proceedings). 
Periods have been obtained by the phase dispersion
method (PDM). Important aspects of this study include:
   \begin{itemize}
   \item Study of red variables  -- mainly SR; 188
         variables in the LMC, 35 in the SMC 
   \item Study of eclipsing binaries; 69 EB in the  LMC, 136 in the SMC
   \item Study of Cepheids; 78 in the LMC, 38 in the  SMC
   \item Study of blue variables (many detected); may be Be stars, at least 
         one is a so-called bumper, resembling a microlensing light curve
   \end{itemize}
   
   By using another selection, 282 AGB variables were found in the LMC
   and 69 in the SMC (see Takeuti et al. in these proceedings). The
   periods and the $K$-band magnitudes of most of these stars are
   comparable with those in the literaure.  The relation between the
   period and the mean colour for the AGB variables is reported.
   
   Another study has been by N.\ Rattenbury (Auckland Univ., MSc), using
   selected MOA fields for series 1 and 2 data. Variable stars were
   detected by deriving the Welch-Stetson variability index (Stetson,
   1996; Welch \& Stetson, 1993).  Only about 2\% of the stars with
   greatest variability were selected for further study in this
   preliminary survey.  Typically these stars have a range in the $R$
   magnitude exceeding $12\sigma$.

 Periods were analysed in several ways, the most successful algorithm being
 the  date compensated discrete Fourier transform (DCDFT) algorithm
 of Ferraz-Mello (1981). The following summarizes the preliminary results.

\begin{figure}[t]
\plotfiddle{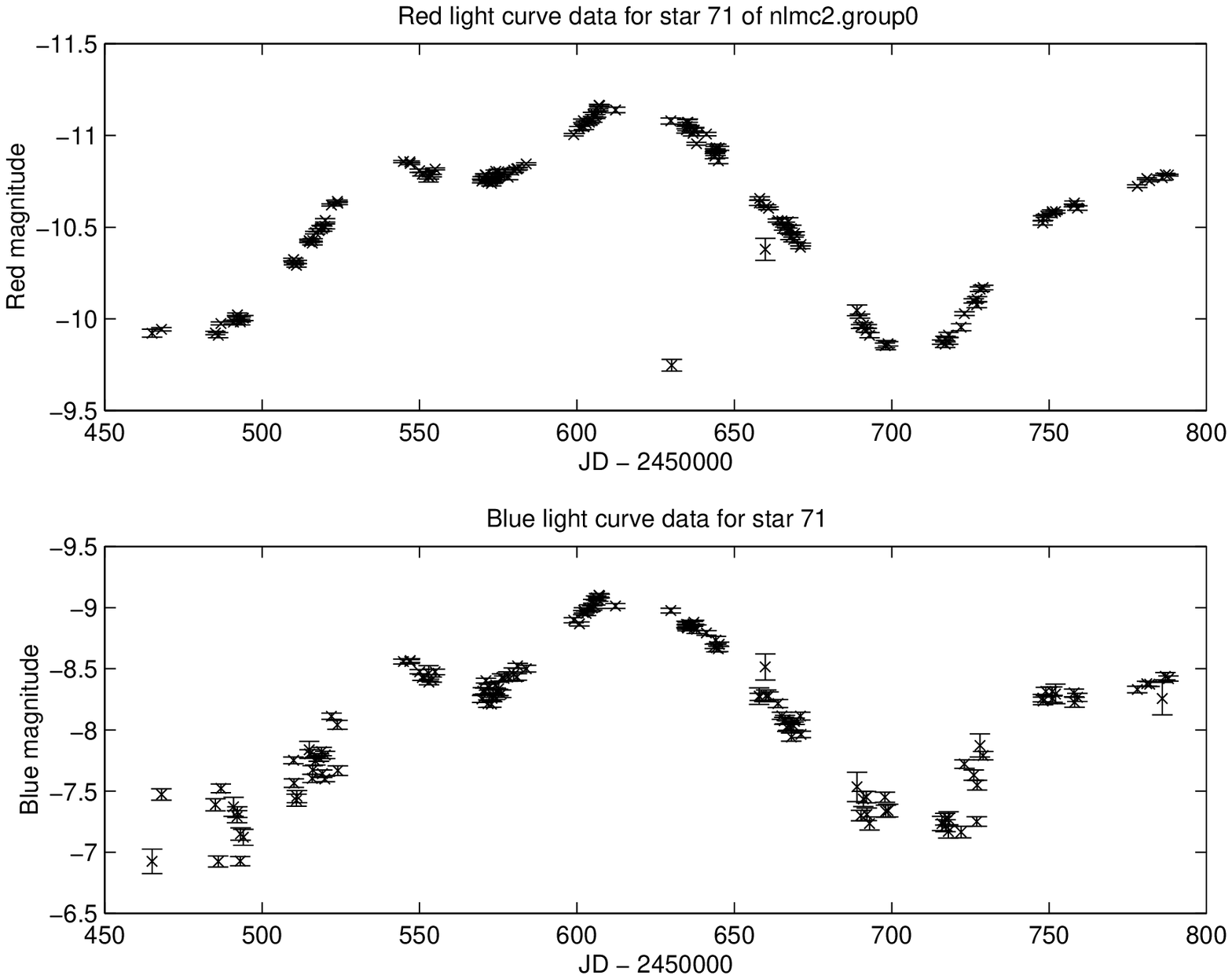}{9cm}{0}{40}{40}{-230}{30}
\plotfiddle{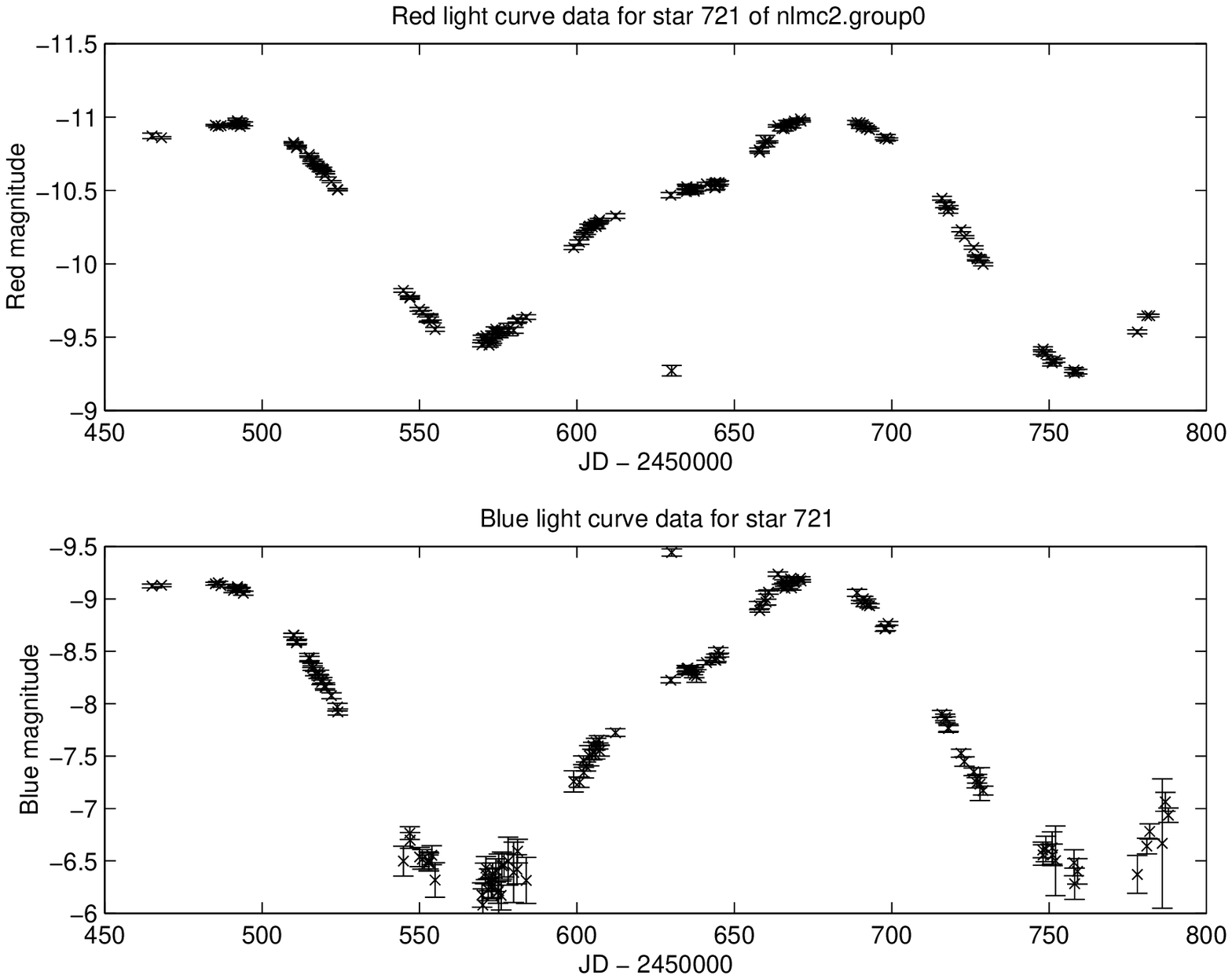}{0cm}{0}{40}{40}{-30}{55}
\plotfiddle{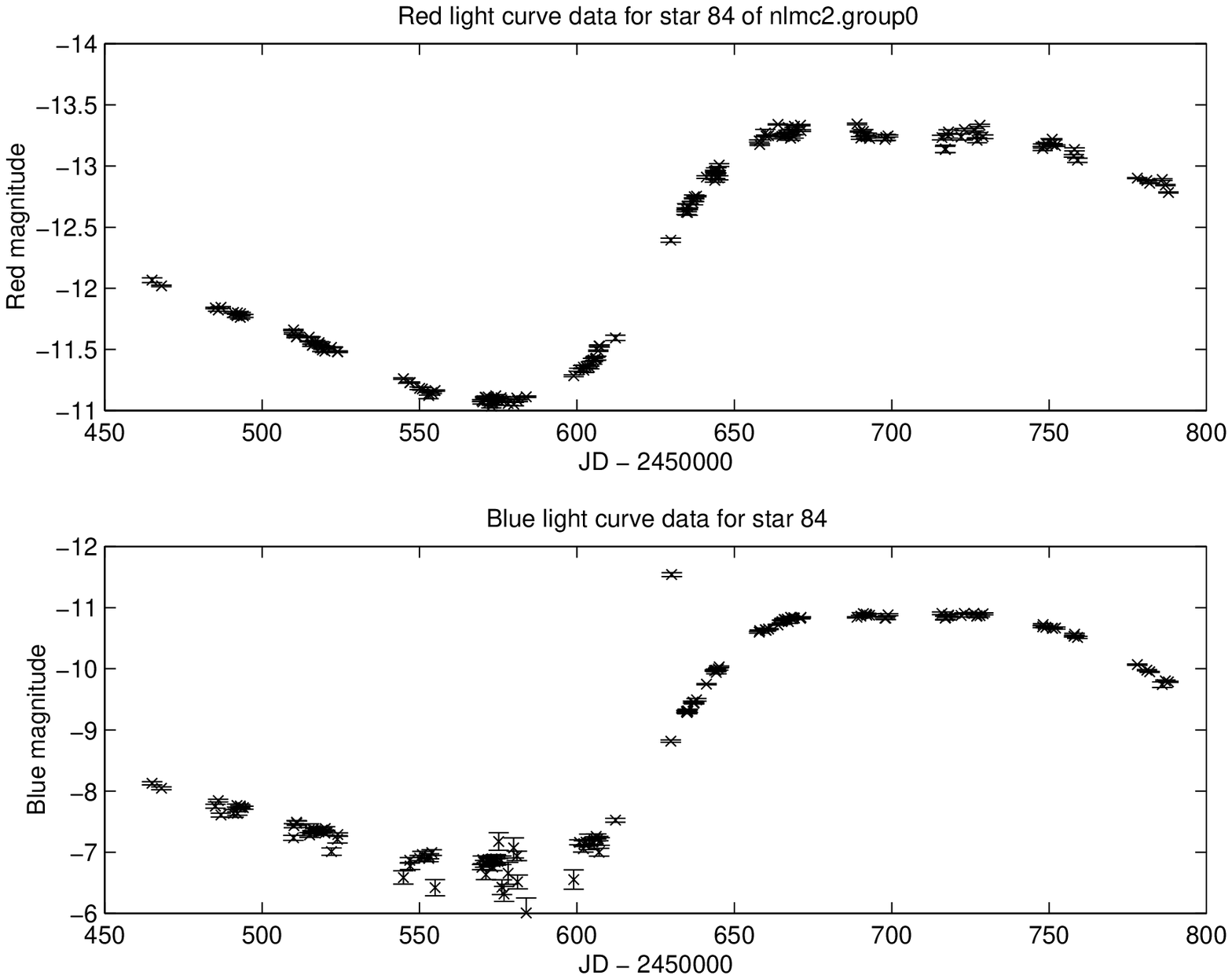}{0cm}{0}{40}{40}{-230}{-85}
\plotfiddle{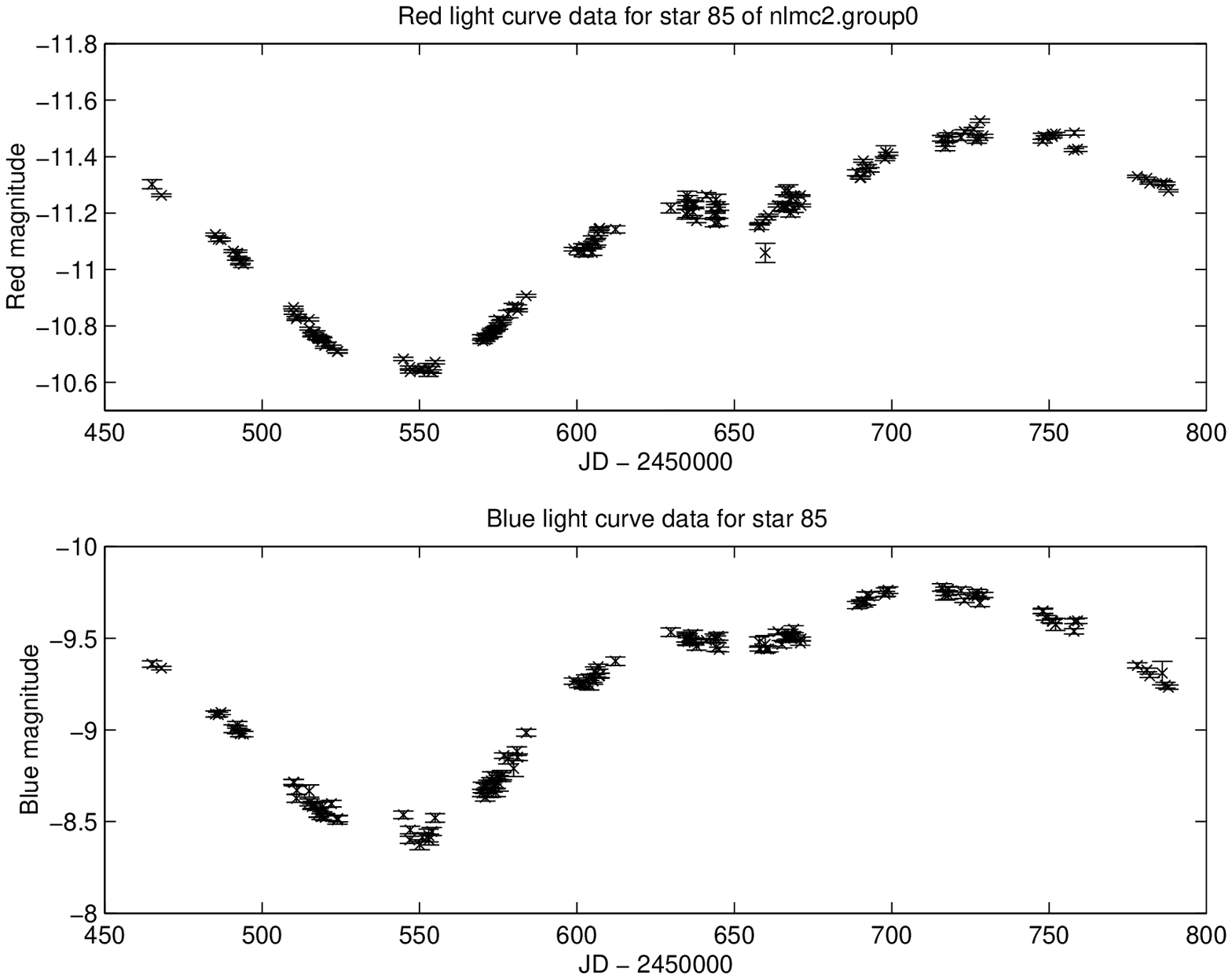}{0cm}{0}{40}{40}{-30}{-60}


\caption{$B$ and $R$ light curves for four long-period red variables in
  field nlmc2}
\end{figure}

\begin{itemize}
\item Study of Cepheids (periods 1--10 d): 70 were identified in the
  LMC and 420 in the SMC
\item Study of Cepheids or semi-regular variables ($P$ 10--50 d): 433 in 
  the LMC, several more in the SMC
\item Eclipsing binaries and RR Lyrae stars ($P$ 1--20 h): 7 in LMC, 71
  in SMC
\item Blue variables ($(B-V)_{\rm J}<0.36$): 37 were identified in the 
  SMC including 7 of long or irregular period, 2 possible ``bumpers'', 
  and at least 3 eclipsing binaries.         
\item  Red variables with $(B-V)_{\rm J}>1.65$ and $P>100$ d 
            (LPV, SR): 89 were found in the SMC and 5 in the LMC
\end{itemize}

Fig.\ 2 shows $B$ and $R$ phased light curves for four of the Cepheids in
the field smc2 (series 2). Fig.\ 3 shows unphased quasi-periodic light curves
for four long period variables from field nlmc2 (series 2).

Finally G.\ Bayne, W.\ Tobin and J.\ Pritchard (Univ.\ Canterbury,
N.Z.) are searching the MOA database for eclipsing binaries and
analysing the light curves (series 1, 2). This work is still in
progress, using the period-searching algorithm of Grison (1994).

\end{document}